\documentclass[11pt]{article}
\usepackage{graphicx}
\usepackage[margin=1.25in]{geometry}
\usepackage[usenames,dvipsnames]{color}
\usepackage{url}
\usepackage[colorlinks = true,
            linkcolor = blue,
            urlcolor  = blue,
            citecolor = blue,
            anchorcolor = blue]{hyperref}

\usepackage{lineno,hyperref,color}
\modulolinenumbers[5]
\usepackage{todonotes}
\usepackage{ulem}
\usepackage{booktabs}
\usepackage{float}
\usepackage{wrapfig}

\usepackage{appendix}
\usepackage{pdfpages}
\usepackage{hyperref}
\usepackage{geometry}                
\usepackage[parfill]{parskip}    
\usepackage{graphicx, subfigure}
\usepackage{amssymb}
\usepackage{amsmath}
\usepackage{epstopdf}
\usepackage{array}
\usepackage{lineno}
\usepackage{scrextend}

\def\Title#1{\begin{center} {\LARGE #1 } \end{center}}
\def\Author#1{\begin{center}{ \sc #1} \end{center}}

\newenvironment{Abstract}{\begin{quotation} \begin{center}
                       ABSTRACT
     \end{center}\bigskip  }{\end{quotation}}

\def\beq{\begin{equation}}
\def\eeq#1{\label{#1}\end{equation}}
\def\eeqn{\end{equation}}

\newenvironment{Eqnarray}%
   {\arraycolsep 0.14em\begin{eqnarray}}{\end{eqnarray}}
\def\beqa{\begin{Eqnarray}}
\def\eeqa#1{\label{#1}\end{Eqnarray}}
\def\eeqan{\end{Eqnarray}}

\def\overbar#1{\overline{#1}}

\let\bar=\overbar

\def\lsim{\mathrel{\raise.3ex\hbox{$<$\kern-.75em\lower1ex\hbox{$\sim$}}}}
\def\gsim{\mathrel{\raise.3ex\hbox{$>$\kern-.75em\lower1ex\hbox{$\sim$}}}}

\def\O{{\cal O}}

\def\del{\partial}
\def\Dslash{\not{\hbox{\kern-4pt $D$}}}
\def\dslash{\not{\hbox{\kern-2pt $\del$}}}
\def\pslash{\not{\hbox{\kern-2pt $p$}}}
\def\ETmiss{\not{\hbox{\kern-4pt $E$}}_T}

\def\Dlr{\mathrel{\raise1.5ex\hbox{$\leftrightarrow$\kern-1em\lower1.5ex\hbox{$D$}}}}

\def\MSB{{\bar{M \kern -2pt S}}}
\def\msb{{\bar{\scriptsize M \kern -1pt S}}}

\def\drb{{\bar{\scriptsize D \kern -1pt R}}}

\input{defs}

\newcommand\snowmass{\begin{center}\rule[-0.2in]{\hsize}{0.01in}\\\rule{\hsize}{0.01in}\\
\vskip 0.1in Submitted to the  Proceedings of the US Community Study\\ 
on the Future of Particle Physics (Snowmass 2021)\\ 
\rule{\hsize}{0.01in}\\\rule[+0.2in]{\hsize}{0.01in} \end{center}}

\begin{document}


\Title{Higgs Factory Considerations}

\bigskip

\Author{
J.A.~Bagger$^{1}$, 
B.C.~Barish$^{2,3}$,
S. Belomestnykh$^{4}$, 
P.C.~Bhat$^{4}$, 
J.E.~Brau$^{5}$, 
M.~Demarteau$^6$,
D.~Denisov$^{7}$,
S.C.~Eno$^{8}$, 
C.G.R.~Geddes$^{9}$,
P.D.~Grannis$^{10}$, 
A.~Hutton$^{11}$,
A.J.~Lankford$^{12}$,
M.U.~Liepe$^{13}$, 
D.B.~MacFarlane$^{14}$,
T.~Markiewicz$^{14}$,
H.E.~Montgomery$^{11}$, 
J.R.~Patterson$^{13}$, 
M.~Perelstein$^{13}$, 
M.E.~Peskin$^{14}$, 
M.C.~Ross$^{14}$, 
J.~Strube$^{15,5}$, 
A.P.~White$^{16}$,
G.W.~Wilson$^{17}$}

\begin{center}
$^{1}$Johns~Hopkins~University,
$^{2}$Caltech, 
$^{3}$U.C.~Riverside, 
$^{4}$Fermilab, 
$^{5}$University~of~Oregon,
$^{6}$ORNL, 
$^{7}$Brookhaven~Lab,
$^{8}$University~of~Maryland,
$^{9}$LBNL,
$^{10}$Stony~Brook~University,
$^{11}$Jefferson~Lab,
$^{12}$U.C.~Irvine, 
$^{13}$Cornell~University, 
$^{14}$SLAC,
$^{15}$PNNL,
$^{16}$University~of~Texas,~Arlington,
$^{17}$University~of~Kansas
\end{center}


\medskip
\begin{center}
\today
\end{center}

\medskip

 \begin{Abstract}
\noindent 
We discuss considerations that can be used to formulate recommendations
for initiating a lepton collider project that would provide precision studies of the Higgs boson 
and related electroweak phenomena.
\end{Abstract}

\snowmass

\def\thefootnote{\fnsymbol{footnote}}
\setcounter{footnote}{0}
%


\section{Introduction}
The International Committee on Future Accelerators (ICFA)~\cite{ICFA_history} has called for the construction of an $e^+e^-$ collider as the next global accelerator project after the LHC~\cite{ICFA}.   
Such a collider would complement the discoveries of the LHC by carrying out precision measurements of the heaviest Standard Model (SM) particles and  discovering potential new phenomena, so as to further our understanding of fundamental physics.  

This call was repeated and amplified in the 2020 update of the European Strategy for Particle Physics~\cite{EPPSU}, which  states ``An electron-positron Higgs factory is the highest-priority next collider.”  

During the Snowmass 2021 process, a set of potential lepton colliders that would operate in the energy region from the $Z$ boson mass to the TeV scale was considered by the Implementation Task Force (ITF) of the Accelerator Frontier group~\cite{ITF} and the relevant parameters were collected.   
In parallel, the Snowmass Energy Frontier group has examined the physics goals to be addressed.
 These colliders have a common goal of producing large samples of Higgs bosons, although they can also be operated to target other physics goals.

In this White Paper, written by members of the Americas Linear Collider Committee (ALCC), we take the ITF list and develop considerations that could be used to make a recommendation on the optimum strategy for realizing a future lepton collider.
Our many years of engagement with the technologies and physics requirements of these colliders provides the basis for our undertaking this report. 
Despite the ALCC emphasis on the International Linear Collider (ILC) and the Compact Linear Collider (CLIC) over the past several years, we are motivated by the desire to investigate the physics associated with the Higgs boson and related precision tests, for whichever lepton collider can be realized.
The criteria for a recommendation include physics requirements, collider design characteristics, and more general considerations.  
They also imply questions of siting such a machine: we believe that it is appropriate to consider whether a Higgs factory might be built in the US with international cooperation.   

We refrain from making a recommendation for the optimal route to a Higgs factory.  Such a recommendation is expected to be made by the Particle Physics Project Prioritization Panel (P5), to be formed subsequent to the report of the Snowmass 2021 workshop.  Thus, this paper is written to offer help for the Snowmass evaluation of Higgs factories in its report and in the anticipated P5 recommendations.

In Section 2 we summarize the history of Higgs collider activity over the past several years.
Section 3 lists the potential projects considered by the ITF.
In Section 4  we identify a set of 
considerations based on the physics goals of a lepton collider.   
In Section 5 we discuss the technical considerations for the collider projects themselves and continue
in Section  6 with more general considerations.
In each of these three Sections we first identify a set of considerations which we think are the most important.  
These are followed by additional criteria, listed without prioritization, that could be taken into account.
These objective considerations can clarify the situation but, perhaps surprisingly, will not lead to a clear conclusion.
We thus find it useful to discuss two further topics. 
Section 7 addresses the role that the United States could serve, either as a host for a Higgs Factory or as participant  in a project elsewhere.
Section 8 discusses the potential evolution of future facilities, the consideration of costs and  the need for global coherence.

\section{History}
In 2013, ICFA formed the Linear Collider Collaboration (LCC) ``to coordinate global research and development efforts for  next-generation particle physics colliders''.   
The Linear Collider Board (LCB), comprised of members from Asia, Europe and the Americas, was charged by ICFA to oversee the LCC.
The Americas Linear Collider Committee was subsequently formed to provide a liaison for physicists and funding agencies in the US and Canada with the LCC/LCB and the global communities.  
At the time ALCC was formed the only colliders on the horizon were linear (the ILC and CLIC) rather than circular, hence the ALCC name.  
The most technologically advanced possibility was the ILC whose technical design report was completed in 2013~\cite{ILC_exec}.   
Physicists in Japan proposed that the ILC be sited in Japan with support from nations in Europe, the Americas and Asia.
The LCC focused much of its attention on R\&D on remaining topics for ILC, developing plans for a proposed site and attempting to obtain Japanese government approval for the ILC.

Meanwhile the CLIC design based on RF accelerating gradients derived from deceleration of a high current drive beam was furthered through a series of subsystem tests.  These now establish the technical feasibility for a first implementation as a 380 GeV collider on a technically limited schedule of the mid 2030's~\cite{CLIC_progress}. 

In 2020, the LCC reached the end of its mandate.  
Given the prospect of Japanese hosting of the ILC, ICFA created the International Development Team (IDT), whose role was to make preparations for the ILC Pre-Laboratory in Japan. 
The mission of the Pre-Laboratory~\cite{ICFA_IDT} would be to finalize the ILC engineering design and to define the organization and global funding arrangements.  
However, support for the ILC within the Japanese government and the timeline for creation of the Pre-Laboratory remain uncertain.

Based on the very successful CERN experience of the sequential LEP and LHC colliders in the same tunnel, new circular $\epem$ Higgs factories have been proposed with sufficiently large circumference to keep radiative losses for $e^\pm$ beams manageable, and to allow 100-TeV scale $pp$ colliders with achievable magnet technologies. 
The recent European Study Update (ESPPU) favored a strategy in which a repeat of the LEP/LHC pattern would result in a Future Circular Collider sequence (FCC-ee, FCC-hh)  following after the HL-LHC. 
This recommendation was accepted by the CERN Council and a FCC Feasibility Study, to report to the next ESPP in four to five years, is under way. 
A similar approach has been proposed in China with the sequential CEPC $\epem$ 
  and SppC $pp$ colliders.

Work toward a muon collider that could operate at the Higgs resonance as well as at higher energies intensified in the past year and the CERN-led International Muon Collider Collaboration has been formed.
 
 Given the long-stated goals of ICFA and the increasing appreciation throughout the world of the importance of precision studies of the Higgs boson, it is natural now to consider the broad range of facilities and possible sites to do this physics.

\section{Potential Higgs factory projects}

The candidate projects considered by the ITF are:

\noindent
{\bf A.}	Conventional circular $e^+e^-$ colliders, with follow-on 100 TeV-class $hh$ colliders
	\begin{itemize}
	    \item FCC-ee (at CERN)~\cite{FCCee,FCCee_status}
	    \item CEPC (in China)~\cite{CEPC}
	\end{itemize}

\medskip

\noindent
{\bf B.}	Linear \epem colliders of varying degrees of maturity and ambitions to reach TeV-scale energies
	\begin{itemize}
	    \item ILC~\cite{ILC_WP}
	    \item CLIC~\cite{CLIC}
	    \item C$^3$~\cite{CCC,XCC}
	\end{itemize}
	
\newpage

\noindent
 {\bf C}.	 Energy recovery has been demonstrated in linacs, including examples with superconducting RF of modest energy scale. 
This group of colliders includes both circular and linear energy-recovery machines. A recent European Roadmap for Accelerator R\&D~\cite{Hutton-panel} deemed them to be considerably less mature than their more conventional siblings.  
    \begin{itemize}
        \item 	CERC  (circular FCC-like collider with energy recovery)~\cite{CERC}
        \item ReLiC (Opposing linacs that alternately accelerate and decelerate electrons and positrons)~\cite{ReLiC}
        \item ERLC (two parallel linacs in each side with the energy from decelerating one beam feeding the beam that is accelerating)~\cite{ERLC}
    \end{itemize}

\medskip

\noindent
{\bf D.} Two Fermilab site filler Higgs factories at about 250 GeV have been motivated by a desire to reawaken interest in a potential US re-entry into the collider field.
  \begin{itemize}
  \item 	Circular $\epem$ FNAL site filler~\cite{FNAL_circ}
  \item Linear $e^+e^-$ FNAL site filler~\cite{FNAL_lin}
 \end{itemize}

\medskip

\noindent
{\bf E.} All the machines above are based on electron-positron collisions. The alternative of a muon collider is the sole member of this group. Interest in such a machine has waxed and waned. Its pursuit was shelved by the 2014 P5 Report but was recently re-established as a possible future direction in the ESPPU report. 
   \begin{itemize}
    \item	Muon Collider~\cite{MuonColl_WP,MuonColl}
    \end{itemize}

\bigskip

\section{Physics considerations}
A set of physics questions for study at lepton colliders operating in the energy range from the $Z$-pole to the TeV region is being discussed in detail by the Energy Frontier group for Snowmass 2021.  
These will provide the primary scientific guidelines for making recommendations on a particular facility.  It is important to note that the scientific questions to be addressed at a lepton collider are quite well defined; the discoveries of the recent hadron colliders have provided clear targets of opportunity at known energies of operation. 
The physics program is not a green field exploration for most of its targets.
 
We believe that the questions that must be addressed with the highest priority are: 

{\it P1. Precision measurement of Higgs couplings to SM fermions and gauge bosons}
\newline 
A Higgs factory should make precision measurements of the couplings and total width of the Higgs boson. Their measurement, compared with the precisely predicted Standard Model values, gives an incisive way to seek new physics and provides a strong indication of its nature.

{\it P2. Measurement  of Higgs self-couplings}
\newline
The Higgs self-coupling establishes the form of the Higgs potential.  Any deviation from the Standard Model form is an incontrovertible sign of new physics.

\newpage
{\it P3. Sensitivity to rare or non-SM Higgs decays}
\newline
Direct measurement of Higgs decays to invisible particles or other exotic final states will help pin down the nature of new physics. 
The ability to measure the total width at a lepton collider offers another avenue to sense new physics.   

All lepton colliders considered by the ITF do address the above criteria associated with the Higgs boson properties, but differ in their capabilities for exploring other related questions. 
There are thus additional physics criteria which differentiate candidate facilities, based upon their ability to study other massive particles
and to infer the presence of new physics beyond the SM.  Among these other criteria are:

{\it P4. Discovery potential for new non-SM physics}

{\it P5. Ability to directly measure top electroweak and Yukawa couplings}

{\it P6. Sensitivity to new physics through precision measurement of loop effects}

~~~~~{\it P6a. Precision top mass and width measurements}

~~~~~{\it P6b. Improvement of precision of $Z$-pole parameters}

~~~~~{\it P6c. Improved $W$ mass measurement}

{\it P7. Ability to improve precision of the strong coupling constant}


\section{Considerations relating to Higgs factory technical issues}
Each of the Higgs factories discussed in the ITF report is capable of operation below and/or above the energy for Higgs production, although sometimes with substantial modification.  
Possible further-future extensions of the energy reach of linear collider facilities to many TeV include those using the very high accelerating gradients provided by laser or beam-driven plasmas or structure wake fields to extend energy reach to many TeV.

We judge the highest priority considerations to be:

{\it T1. Range of possible operating energies and ease of changing energy}
\newline 
Ideally, a new facility would be capable of addressing physics questions over a range of energies.

{\it T2. Annual integrated luminosity recorded by all experiments}
\newline 
To the extent that the precision of measurements is controlled by statistical uncertainty, the integrated luminosity delivered to the experiments is of prime importance.

{\it T3. Upgradability to higher energy and luminosity}
\newline 
Searches for new particles or deviations from the SM-predicted precision measurements are enhanced by operating at energies in the TeV range or by increases in luminosity.

\newpage
{\it T4. Extent and cost of remaining R\&D}
\newline 
The extent to which a proposed facility is ready for approval and construction start can be measured by the number and criticality of needed R\&D projects still remaining, their estimated cost and, importantly, their likely duration.  In some cases the R\&D may be absolutely critical to confirm feasibility.


More specific technical criteria which could differentiate among potential projects include: 

{\it T5. Ability to operate at the top pair threshold}

{\it T6. Ability to run at the $Z$ pole}

{\it T7. Ability to run at the $WW$ threshold}

{\it T8. Collision energy stability and calibration precision at all operating energies }

{\it T9. Beam position stability and luminosity calibration precision}

{\it T10. Ability to control beam-related backgrounds at the IR}

{\it T11. Ability to provide independent confirmations of new discoveries}

{\it T12. Ability to provide polarized electrons}

{\it T13. Ability to provide polarized positrons}

{\it T14. Possibility to reconfigure as a $e^-e^-, e^-\gamma$ or $\gamma \gamma$ collider}

{\it T15. Possibility to reconfigure as a $hh$ or $ep$ collider}

{\it T16. Opportunities for beam dump experiments, long lived particles, etc.}

{\it T17. Need for, and scientific utility of, technology demonstrators}

\bigskip

\section{Considerations relating to  more general project issues}

The final category of criteria for developing recommendations is more general than those in the preceeding sections, but is perhaps even more important.   The most important of these in our view are:

{\it G1. Construction cost}
\newline 
The construction cost of any of the accelerator and detector facilities under consideration is sufficiently large as to require international collaboration, and will be the dominant factor in securing agreement to undertake a project.

{\it G2. Possible start date of physics}
\newline 
From the physics point of view it is of course desirable to start a new collider as soon as possible. 
The start date is controlled to a large extent by the cost and the suite of major scientific projects already planned or in construction by the Host nation or region.  
For example, the FCC-ee at CERN is constrained by the operation of the HL-LHC.  
Funding for a new collider in the US would be constrained by the construction of DUNE/LBNF.  
In the US, the timing would also depend on the sequence of major projects from the several offices within the DOE Office of Science.
The need to coordinate the timing of the funds from many participating nations may introduce further constraints.
The time required to conduct needed R\&D (Section 5, item {\it T4}) can also affect the start date.

{\it G3. Sustainability and operating costs}
\newline 
The environmental sustainability and cost to operate the facility are major considerations for the acceptance of a new collider.
This need was highlighted in the EPPSU~\cite{EPPSU} and amplified in a study on sustainability for the ILC in Japan~\cite{ILC-sustain}.
 
{\it G4. Suitability as basis for follow-on facilities}
\newline 
Many projects in the past have been made possible by the existence of infrastructure built for a previous facility.  The nature of scientific research tells us that the discoveries of today will dictate the needs for future research.   And it is almost always the case that the next steps will be more extensive and expensive than those that came before.  So the effective re-use of prior infrastructure is essential.

{\it G5. Broader impacts on society through technological innovations}
\newline 
Although fundamental scientific questions motivate the new facilities, governments also gauge their desirability on the basis of benefits to the wider society.   New technologies with wider capabilities that are enabled by prospective scientific projects are a strong selling point for their approval.

Other general considerations for recommendations include:

{\it G6. Compactness, extent of surface disruption, ability to obtain land use agreements}

{\it G7. Environmental and radiation concerns}

{\it G8. Synergistic use of a facility for other science/engineering studies}

\section{Domestic US Considerations}
For almost a decade, the US focus has been on the ILC in Japan, based upon the recognition that it is the most technically advanced, and that Europe and the US presently have compelling near-term priorities.  
The situation has now changed, with the lack of progress in securing government support for ILC in Japan, the maturation of drive-beam RF for a linear collider at CERN, and the development of attractive circular $e^+e^-$ collider proposals in Europe and China.  
In this more fluid situation, it will be important for the US to work toward finding a solution that leads to the execution of at least one of the projects somewhere in the world.   
As the “Americas” Linear Collider Committee, it would be negligent not to discuss the possibility of US hosting.  
In this section, we point out the positive aspects, although we recognize the hurdles to be surmounted in realizing a US-hosted facility.

\subsection{US contributions to accelerator projects}
The US has a long record of technical advances in accelerator technology that positions it well to lead a project either at home or elsewhere in the world.  
The US expertise in accelerator science and in building large accelerator facilities is longstanding and deep, with major projects undertaken in the DOE offices of High Energy Physics, Nuclear Physics, and Basic Energy Sciences. 
The Global Design Effort that led to the technical design report for the ILC was based on the leadership by US physicists and engineers in many areas of accelerator science and technology.
This experience, and the flexibility shown by members of the US community, gives an excellent springboard for leadership of a new Higgs factory collider.

 The core technology for the ILC, superconducting radio frequency (SRF) acceleration, has matured and expanded in the US and worldwide.  
 The circular colliders considered in the ITF report would also use SRF cavities.
 CEBAF at Jefferson Lab upgraded its energy threefold over its design due to its use of SRF acceleration.  
 The DOE Office of Science has invested in such infrastructure at its labs, enabling the construction of SNS at Oak Ridge, LCLS-II and LCLS-II-HE at SLAC, FRIB at Michigan State University, and now PIP-II at Fermilab. 
 The Electron Ion Collider project currently in preparation at BNL relies, like other modern circular machines, on SRF acceleration.  
 
SLAC has been a world leader of high-gradient normal-conducting RF acceleration for decades.  
The SLAC linac, the SLC, the FACET test facility and the engagement in CLIC R\&D have recently led to the development of the C$^3$ (Cool Copper Collider) proposal for a Higgs factory.  
Variations of this technology are used at Paul Scherrer Institute (SwissFEL),  Trieste (FERMI) and SLAC (LCLS) free electron lasers, and are leading to the development of compact accelerators for a variety of uses. 
The path to maturity for CLIC, and even more so for C$^3$, is longer than that for the ILC, but there has been considerable excitement within the accelerator community for these technologies. These options have the potential for increased accelerating gradients over the SRF option, hence a reduced linear collider footprint and potential cost savings.  

The US has a history of pioneering work in circular accelerator construction at the largest scales with the Tevatron, RHIC and the LHC.   
The development of powerful magnets for circular colliders that was pioneered in the US was key contributions to the LHC.  
The realization of energy-recovery techniques has been refined and applied in US projects.   
The US expertise in beam dynamics has enabled the physics programs at CESR, PEP-II, the Tevatron, the Main Injector, and SLC.    Much of the development of a muon collider has been accomplished by the US-based Muon Accelerator Project (MAP)~\cite{MAP}.

It is easy to see how the US could play a lead role, at home or abroad, in a linear or circular $e^+e^-$ collider based on either superconducting or normal conducting RF.     

In its previous incarnation, P5 made the recommendation to shelve the Muon Collider efforts. 
They were judged to be diverting the community and required large investments that were difficult to provide given the commitments to the LHC and those emerging for the neutrino program. 
Nevertheless, we note that the European community, having raised the question of a Muon Collider and the idea to contemplate a demonstrator design for the next ESPPU, is looking to the US for experienced help.

\subsection{US contributions to collider detectors}
The US has been at the forefront in building and operating detectors for colliders for decades.   
The detectors at hadron or $ep$ colliders at the ISR, Tevatron, HERA and LHC were built with the leadership of US physicists, covering all of the major subsystems.   
Detectors at $e^+e^-$ colliders have evolved from those at SPEAR, PEP and CESR, through those at SLC, LEP, KEK-B and PEP-II, and now on to the Higgs factories.    
US physicists were the spokespersons or top-level detector and physics leaders in most of these experiments.

In the past decade or more, US physicists have taken lead roles in the designs of both of the ILC detectors ~\cite{ILC_det}, and have played the dominant role in the SiD consortium.  
Innovative concepts for lepton collider detectors continue to surface and although the funding for lepton collider detector R\&D in the US has been minimal over the past decade, the depth of expertise in the US community is great.   The ILC detector designs have been taken over with modest modifications to detectors being proposed at CLIC, FCC-ee and CEPC.

US physicists have led the world-wide articulation of the lepton collider physics program, and have contributed significant new ideas that will broaden its scope.   Many of the software tools developed for the study of lepton collider physics processes and the collider backgrounds have been initiated by the US community.

There can be no doubt that US physicists will be important leaders for the physics and detector programs at a lepton collider hosted either domestically or abroad.

\subsection{R\&D needs}
Test facilities at ANL, BNL, Cornell, FNAL, LBNL and SLAC offer a wide range of opportunities for R\&D on new methods and technologies for accelerators. 
These facilities will promote advances in plasma and dielectric structure wakefield acceleration, new RF techniques, and development and test of new SRF technology. 
Their operation fosters the training of accelerator scientists over a broad range of topics, giving the basis for undertaking major new projects.

Whatever choice is made for a future Higgs factory, we believe that the US should be a leading participant. 
This implies the need for substantial R\&D for technical areas (including detectors) to which the US could contribute, and also for demonstrator projects for new collider concepts or major subsystems. 
In the case of some proposed facilities, much of this R\&D has already been done, but for others, most remains. 
For the ILC, a first demonstrator project was the European XFEL, and a second is LCLS-II. 
The choice of scale for a demonstrator needs to balance the desirability of having a machine which is useful scientifically with the desire to invest as minimally as possible in this R\&D phase.  
A general rule of thumb for such R\&D funding is about 10\% of the total project cost.  
A discussion of the need for enhanced funding for Higgs factory R\&D can be found in a White Paper submitted for Snowmass 2021~\cite{HiggsFactRD}.

\section{Global considerations}

\subsection{Evolution of facilities}
As indicated by criterion {\it G4} above, a new project that lends itself naturally to extension to further facilities is attractive in creating a basis for the present while enabling a future path in a relatively cost-effective manner.   
Many of the possible Higgs factories mentioned in Section 3 could be imagined to allow new opportunities through follow-on upgrades for many decades.  

The circular colliders are based on a large tunnel, sized for a very high energy hadron-hadron collider. 
The tunnel size is dictated by the goal of 100 TeV $pp$ collisions using foreseeable magnet technology.   
However, the hadron machine requires magnets  that do not currently exist and will require considerable further R\&D.
The hadron-hadron and electron-positron circular colliders are linked in several ways. 
The large tunnel needed for the hadron collider facilitates the $e^+e^-$ collider with technology that is relatively well proven. 
But since the full energy hadron machine needs major advances in magnet technology, an $e^+e^-$ precursor is natural for filling the time gap that may occur in achieving them, thus maximizing the physics exploitation of the tunnel.

A linear collider Higgs factory can be extended in length, or its acceleration gradient upgraded, to increase the energy.  
In these scenarios, the tunnel is a major reusable component that can be filled with quite different RF technology from that used in the initial incarnation.
Advances in SRF offer energy extension up to 3 TeV or beyond~\cite{Padamsee}.  A change at some point to a CLIC or C$^3$ technology might also provide a 3 TeV horizon.  
If very high gradient plasma acceleration can be developed as a viable alternative; the linear tunnels could offer the site for even further improvements in collision energy.

The ERL-based Higgs factories offer a longer term evolution in energy and luminosity that could be achieved through upgrades based on either the circular or linear collider infrastructures of Groups A or B.
Their relevance for the present discussion depends on the anticipated window to advance the required R\&D. 
Of course, without support, that R\&D will not happen.

Many challenges remain before a muon collider could be realized and a
relatively long period for R\&D is needed. 
Such a facility would necessarily require novel systems for a high-power proton driver, muon capture, cooling and subsequent acceleration.  
It is possible that a first step of a muon-based facility could be a high intensity neutrino factory.  
In any case the front end systems for muon production and capture could be the springboard for later expansion to higher energy.

\subsection{Costs}
The total cost of the facility is the dominating consideration for the multiple nations needed to join together in building a new collider.  
The question of how much cost is acceptable depends on many factors, including the expected scientific impact, the technological
benefits accruing from the project,  the broader societal sense of its importance, and  political considerations of
international cooperative ventures.  
The acceptable cost in the US could also depend on the range of potential partners from within the broader DOE and NSF programs.
If a new facility were to be based in the US with some contributions from other regions, one could look to the scale of 
current or recent projects such as EIC~\cite{EIC}, PIP-II~\cite{PIPII} or LBNF/DUNE~\cite{LBNF_DUNE}.  
The final cost of NASA's James Webb Space Telescope of about \$10B (US accounting), with foreign contributions at the level of around 5\%, represents an upper limit for the cost of a new US-based project that will be difficult for HEP to reach.  
For US participation in projects abroad, the precedents include LHC (\$531M  in FY2008 for the initial accelerator and 
detector construction), 
ITER (US share of 9.1\% of a total cost which has risen to over \$20B), and the expectation that US participation in 
ILC in Japan would require about a contribution of at least 10\%, but perhaps up to 25\%,
 of the full project cost 
of about \$7.5B (Japanese/European accounting) for the initial collider, excluding detectors.  
These comparisons must of course be made with care due to the differences in cost accounting practices in 
different regions, differing definitions of what is included in the cost of different projects, and  fluctuating exchange rates 
(it can be preferable to compare costs in different nations using Purchasing Power Parity~\cite{ILC_tdr}).

\subsection{Global coherence}

Up until fairly recently, comparable new high energy physics facilities have been built in different regions of the world that have often provided independent confirmation of results.  PEP-II, PETRA and TRISTAN all gave similar opportunities.   
Several B-factories (PEP-II, KEK-B and CESR) were built.  
The CERN Sp$\overbar{\rm p}$S and the Fermilab Tevatron, or RHIC and the LHC ion collider plowed similar furrows.  
Only when the projects became as large and costly as the LHC did the world enter the stage where only one facility of a given type could be envisioned.

It is not realistic to imagine that scientific advisory panels alone are sufficient to insist on a rational array of complementary facilities in different parts of the world, thus optimizing the global scientific payoff.  
But it is nevertheless worthwhile for such panels to explain how their recommendations fit into the global picture and to enunciate the goal of developing a coherent world plan. 

\section{Conclusions}

We have attempted in this paper to discuss in an even-handed manner the issues that we imagine would need to be considered when attempting to chart a path forward for Higgs physics. 
Our discussion has been rooted in the experience in HEP over the past 40 years or so. 
We embarked on this report  with the goal of an objective assessment of the relative performance of different machines. 
Such a comparison is complicated by the scope of physics beyond that prescribed as mandatory, which differs for the proposed options. This led us to the necessity of considering aspects such as the possibility of a domestic option and/or the global particle physics roadmap. From the outset, we intended to lay out the considerations without making a choice. Our discussions reinforced that intent. We leave the resolution of the conundrum to the upcoming P5 panel, HEPAP and ICFA.

\section{Acknowledgments}
We are indebted to Thomas Roser and the Accelerator Frontier Implementation Task Force for the careful description of the various colliders appropriate for the study of the Higgs boson.   
The talks and discussions in the  ``Snowmass Agora on Future Colliders'' organized by the Snowmass 2021 Energy and Accelerator Frontier working groups and Fermilab have been an important contribution to our understanding of the various potential projects.   
We gratefully acknowledge the support of the US Department of Energy and the US National Science Foundation.

\bibliographystyle{elsarticle-num}
\bibliography{theBIB}

\end{document}